\documentclass[12pt,preprint]{aastex}

\usepackage{graphicx}
\usepackage{epsfig}
\usepackage{txfonts}
\usepackage{longtable}
\usepackage{lscape}
\usepackage{natbib}

\begin{document}

\title{Scattering polarization in solar flares}
\shorttitle{Scattering polarization in solar flares}
\shortauthors{\v{S}t\v{e}p\'an \& Heinzel}

\author{
Ji\v{r}\'\i\, \v{S}t\v{e}p\'an\altaffilmark{1}
and
Petr Heinzel\altaffilmark{1}
\altaffiltext{1}{Astronomical Institute ASCR, Fri\v{c}ova 298, 251\,65 Ond\v{r}ejov, Czech Republic}
}

\begin{abstract}
There is an ongoing debate about the origin and even the very existence of a high degree of linear polarization of some chromospheric spectral lines observed in solar flares. The standard explanation of these measurements is in terms of the impact polarization caused by non-thermal proton and/or electron beams. In this work, we study the possible role of resonance line polarization due to radiation anisotropy in the inhomogeneous medium of the flare ribbons. We consider a simple two-dimensional model of the flaring chromosphere and we solve self-consistently the non-LTE problem taking into account the role of resonant scattering polarization and of the Hanle effect. Our calculations show that the horizontal plasma inhomogeneities at the boundary of the flare ribbons can lead to a significant radiation anisotropy in the line formation region and, consequently, to a fractional linear polarization of the emergent radiation of the order of several percent. Neglecting the effects of impact polarization, our model can provide a clue for resolving some of the common observational findings, namely: (1) why a high degree of polarization appears mainly at the edges of the flare ribbons; (2) why polarization can also be observed during the gradual phase of a flare; (3) why polarization is mostly radial or tangential. We conclude that the radiation transfer in the realistic multi-dimensional models of solar flares needs to be considered as an essential ingredient for understanding the observed spectral line polarization.
\end{abstract}


\section{Introduction\label{sec:intro}}

One of the long-standing issues in solar physics is the problem of the origin and even the very existence of a high degree of linear polarization of some chromospheric spectral lines observed in solar flares. Over the last few decades, several authors have reported the detection of linear polarization signals in solar flares, with fractional polarization amplitudes at the level of several percent \citep[e.g.,][]{henoux90,fang93,vogt96,xu05,firstova12,henoux13}. The observed spectral lines include Ca\,{\sc ii}\,K, Na\,D$_2$, and the hydrogen lines H$\alpha$ and H$\beta$. This polarization is typically not observed in the center of the flare ribbons but rather at their boundaries with the surrounding chromosphere and it is often found to be perpendicular (i.e., radial) or parallel (i.e., tangential) to the nearest solar limb. The standard interpretation of this polarization is in the terms of impact polarization of the atomic levels due to non-thermal particle beams and/or the neutralizing return currents \citep{karlicky02,stepan07b}, even though polarization can also be observed during the gradual phase of a flare \citep[e.g.,][]{hanaoka04}. On the other hand, other authors have reported no observable linear polarization above the 0.1\,\% level \citep{bianda05}. Up to now, most of the interpretations of the observed polarization have been done either in the so-called last scattering approximation \citep[see][and references therein]{vogt01} or using one-dimensional (1D) models taking into account the effects of radiation transfer \citep{stepan07}. A specific version of impact polarization due to evaporative upflows has been proposed by \citet{fletcher98}.

In this letter, we propose a different perspective on the interpretation of such observations. The chromospheric lines of interest are formed under non-LTE conditions and they are susceptible to resonant scattering polarization \citep[see][]{ll04}. The 1D plane-parallel approximation of the flare ribbons must fail at the ribbon edges where the strong horizontal gradients of the physical conditions are found. From the point of view of the radiative transfer theory, the ribbons are a remarkable example of a multi-dimensional medium where the anisotropy and symmetry breaking of the radiation field may be very significant. This can result in enhanced emission of linearly polarized radiation in spectral lines due to resonance scattering. For a basic investigation of resonance line polarization and the Hanle effect in horizontally inhomogeneous stellar atmospheres see \citet{rms11}.

One of the most familiar manifestations of scattering polarization in the Sun is the so-called second solar spectrum \citep{sss97}. In the vicinity of a solar flare, the cylindrical symmetry of the atmosphere is broken and multi-dimensional calculations are necessary. In this letter, we don't consider the possible role of impact polarization and we solve a two dimensional (2D) radiative transfer problem using a simple model of a flaring atmosphere to study the role of scattering polarization and the Hanle effect on the emergent spectral line polarization.

\section{Formulation of the problem\label{sec:formul}}

\subsection{Resonance scattering polarization\label{ssec:polar}}

We describe the polarization state of radiation by means of the vector of Stokes parameters $(I,Q,U)^{\rm T}$, where $I$ denotes the specific intensity, and $Q$ and $U$ quantify the linear polarization. In our calculation, we do not consider the circular polarization due to the Zeeman effect. Following the usual convention, we define the positive $Q$ direction to be parallel to the nearest solar limb. In this work, we apply the theory of spectral line polarization described in \citet{ll04}. The description of the mean intensity, anisotropy, and symmetry properties of the radiation field at a given point of the atmosphere can be done using the irreducible representation of the radiation field tensors, $J^K_Q$, with $K=0,1,2$, and $Q=-K,\dots,K$. The physical meaning of the individual $J^K_Q$ components becomes apparent from their definition \citep[see Sect.\,5.11 of][]{ll04}. In particular, $J^0_0$ corresponds to the familiar mean intensity of the radiation and it is the only non-zero radiation tensor if the field is isotropic.

In order to find the atomic excitation state at every point within the model atmosphere, the $J^K_Q$ tensors need to be calculated at every such point by solving the radiative transfer equation for the Stokes parameters. The radiation field tensors enter the equations of statistical equilibrium whose solution provides the density matrix of the atomic levels, $\rho^K_Q$, where $K=0,\dots,2J$ and $Q=-K,\dots,K$ for the level with the total angular momentum $J$. In this representation, $\rho^0_0$ is proportional to the population of the level and $K>0$ components contain information on the polarization state of the level.
In our model, only the $K=2$ components (alignment) need to be accounted for, while the $K=1$ components (orientation) are identically zero. It follows that levels with angular momentum $J<1$ can only hold population but not the atomic polarization.
After a self-consistent solution of the non-LTE problem, one can synthesize the emergent polarized spectra \citep{stepan13}.

In the case of a cylindrically symmetric plane-parallel atmosphere, all the coherence components $\rho^K_{Q\neq 0}$ are identically equal to zero and only the $Q=0$ components remain. Such a description is adequate in the case of 1D unmagnetized models of the solar atmosphere in which the emergent polarization is either radial or tangential with respect to the nearest limb. If the cylindrical symmetry is broken due to the presence of an inclined magnetic field and/or due to horizontal inhomogeneities of the thermal structure of the plasma, a general description taking into account all the $\rho^K_Q$ elements is necessary \citep[e.g.,][]{rms11}.

If magnetic field is present in the atmosphere, scattering line polarization can be modified via the Hanle effect \citep[see Sect.\,5.7 of][]{ll04}. The Hanle effect typically leads to rotation of the polarization direction and to decrease of the polarization degree of the emergent radiation. The order of magnitude of the magnetic field strength at which the Hanle effect becomes significant for a particular spectral line can be quantified by the so-called critical Hanle field, $B_{\rm H}=1.137\times 10^{-7}/(t_{\rm life}g_J)$, where $t_{\rm life}$ is the lifetime of an involved atomic level and $g_J$ is the Land\'e factor \citep[e.g.,][]{jtb01}.
For chromospheric lines, $B_{\rm H}$ is usually in the interval from milligauss to few deca-gauss.
If the field is significantly stronger than $B_{\rm H}$ (the so-called Hanle effect saturation regime), the quantum coherence in the reference frame in which the magnetic field is parallel to the quantization axis, is destroyed and the polarization direction of the emitted radiation is either parallel or perpendicular to the plane defined by the magnetic field vector and the photon propagation direction.

\subsection{The model atom\label{ssec:atom}}

Given that the goal of this letter is to point out the general mechanism of a possible creation of the linear polarization in solar flares, we do not particularize the model atom to any specific chemical species. Instead, we choose a generic two-level model atom with a resonant transition at $\lambda_0=5\,000$\,\AA, the Einstein coefficient of the spontaneous emission of $A_{u\ell}=10^8\,{\rm s}^{-1}$, and the atomic weight of hydrogen. The angular momenta of the lower and upper levels are $J_\ell=0$ and $J_u=1$, respectively. The line absorption profile is a Voigt profile with a damping parameter $a=10^{-2}$, constant throughout the atmosphere. We neglect the effect of stimulated emission. We assume the approximation of complete frequency redistribution (CRD).

According to Sect.~\ref{ssec:polar}, the lower atomic level with zero angular momentum can only hold population but it cannot be polarized. The only polarizable level is thus the upper level whose non-zero density matrix elements are the population $\rho^0_0$ and the five complex components $\rho^2_Q$ of the alignment. The Land\'e factor of the upper level is $g_u=1$ and the critical Hanle field of our line is $B_{\rm H}=11$\,G.

\subsection{The model atmosphere\label{ssec:atmosphere}}

We consider a simple isothermal 2D model atmosphere with the kinetic temperature $T=6\,000$\,K representing the solar chromosphere.
It corresponds to a vertical slice perpendicular to the flare ribbon with $z$ being the vertical and $x$ being the horizontal axis (see below).
The atmosphere is exponentially stratified with the number density of atoms given by $n(z)=n_0{\rm e}^{-z/H}$, where $n_0=10^{15}\,{\rm cm}^{-3}$ and $H=75$\,km.
We assume periodic boundary conditions in the $x$-direction. The computational domain extends horizontally from $x=-20$\,Mm to $x=20$\,Mm. At these boundaries, the solution is virtually identical to the corresponding 1D solution for a given vertical column of the isothermal atmosphere. This justifies the use of periodic boundary conditions.
Along the $z$ axis, the model extends from $z=0$\,Mm, which corresponds to the photospheric level where the line is thermalized, to $z=2$\,Mm, which corresponds to the optically thin upper chromosphere.

The thermal collisions are characterized by the collisional destruction probability $\epsilon = C^{\rm th}_{u\ell}/(A_{u\ell}+C^{\rm th}_{u\ell})=10^{-2}$, where $C^{\rm th}_{u\ell}$ is the collisional de-excitation rate due to the thermal collisions.
We do not consider any collisional depolarization in this work.

\begin{figure}[t]
\centering
\includegraphics[width=14.cm]{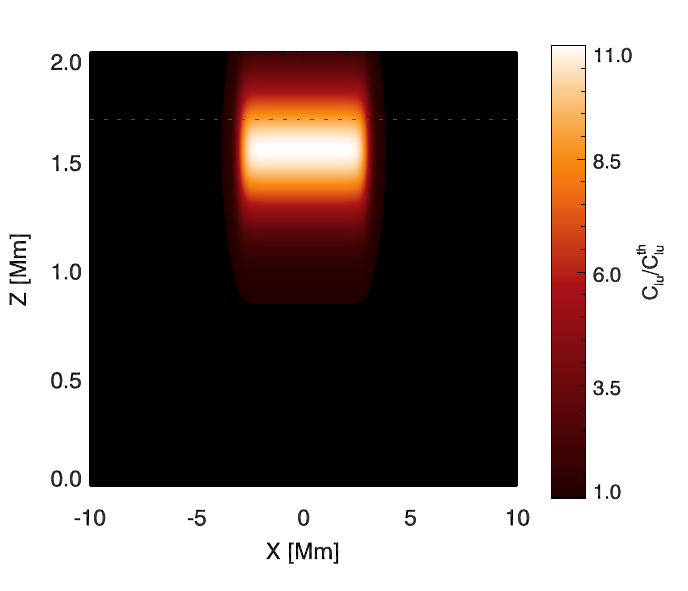}
\caption[]{
Collisional excitation rate $C_{\ell u}/C_{\ell u}^{\rm th}$ within the model atmosphere calculated by Eq.\,(\ref{eq:clu}). The horizontal dotted line at $z=1.69$\,Mm indicates the height where the line-center optical depth is unity for the disk-center observation.
}
\label{fig:colls}
\end{figure}

We model a single flare ribbon by adding the non-thermal collisional excitation rate in the central part of the model atmosphere. According to the theory of non-thermal beam propagation \citep[e.g.,][]{emslie78}, particle beams deposit part of their energy via non-thermal excitation of the atoms. This process is most efficient in the middle and upper chromosphere. In our model, the region of enhanced non-thermal excitation is restricted in both $z$ and $x$ directions in order to mimic the presence of a spatially localized ribbon. Using the sigmoid function $\sigma(w,d,x)=[1+e^{-(x-d)/w}]^{-1}$, we consider the total collisional excitation rate (i.e., thermal plus non-thermal), to be given by an ad hoc expression
\begin{equation}
C_{\ell u}(x,z) = C^{\rm th}_{\ell u} \left[ 1 + 10\, {\rm e}^{-(z-z_0)^2/w_z^2}\sigma(-w_x,d,x)\sigma(w_x,-d,x) \right]\,,
\label{eq:clu}
\end{equation}
where $z_0=1.55$\,Mm is a $z$-coordinate of the maximum of collisional rates, $w_z=0.3$\,Mm controls the vertical extension of the region of enhanced non-thermal collisions, $w_x=3$\,Mm is the half-width of the ribbon in the horizontal direction, and $d=0.15$\,Mm determines the horizontal gradient of collisional rates at the ribbon edges.
The spatial distribution of collisional excitation rates is plotted in Fig\,\ref{fig:colls}.
The above increase of collisional rates is chosen by analogy with more realistic models \citep[e.g.][]{kasparova02,stepan07}.
Note that our choice is rather conservative since the non-thermal collisional rates can exceed the thermal ones by much more than the factor of ten considered here in Eq.\,(\ref{eq:clu}).

Only the collisional de-excitation due to the thermal electrons must be considered while the de-excitation by the high-energy non-thermal collisions can be neglected \citep[e.g.,][]{kasparova02}. The parameter $\epsilon$ is therefore constant in the model atmosphere and the only spatially dependent rate is $C_{\ell u}$.
Note that $C_{\ell u}$ is the population transfer rate and that we do not consider impact polarization in our model.

Magnetic field at the flare-loop footpoints can be assumed to be roughly vertical with a strength of a few hundred gauss. That is well above the Hanle effect saturation of most of the spectral lines of interest and also above the saturation of our generic line. In the following section, we study both non-magnetized and magnetized solutions of the radiative transfer problem.

\section{Results\label{sec:res}}

\begin{figure}[t]
\centering
\includegraphics[width=14.cm]{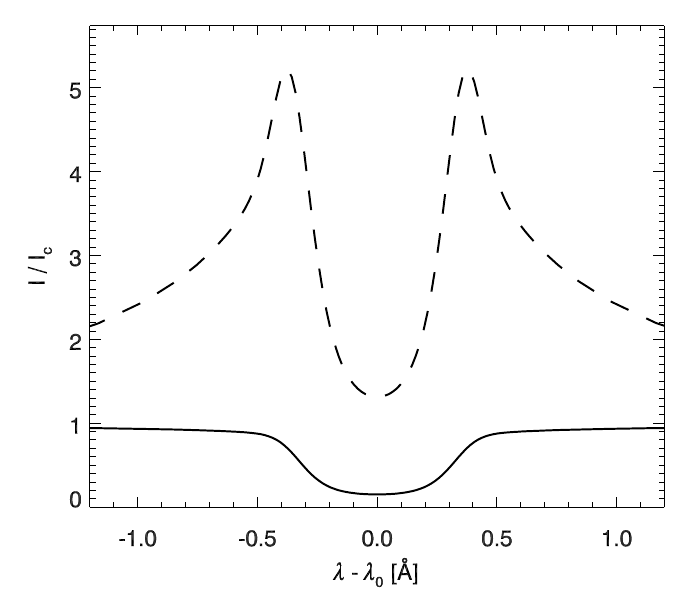}
\caption[]{
Disk-center line intensity profiles normalized to the continuum level for two positions in the model atmosphere.
Solid line: Quiet region ($x=20$\,Mm). Dashed line: The center of the ribbon ($x=0$\,Mm).
The emission in the line wings is due to collisionally increased line source function in the deeper layers of the ribbon atmosphere, i.e., approximatelly at the unit optical depth at the particular wavelength \citep[cf., the line profiles and the contribution functions in Figs.~1 and 5 of][]{kasparova02}. Note that the line Doppler width is $\Delta\lambda_{\rm D}=0.17\,$\AA\/ and we do not consider any overlapping continuum in our semi-infinite atmosphere.
}
\label{fig:int}
\end{figure}

We have solved the non-LTE problem described in the previous section using the radiative transfer code PORTA \citep{stepan13} applied to a 2D grid with $N_x\times N_z = 256\times 100$ grid points.
In Fig.\,\ref{fig:int}, we compare the line intensity profile of the `quiet' region of the atmosphere with the profile obtained at the center of the ribbon. The later resembles the characteristic emission line profiles in solar flares \citep{kasparova02}.

\subsection{Non-magnetized atmosphere\label{ssec:nonmag}}

\begin{figure}[t]
\centering
\begin{tabular}{c}
\includegraphics[width=7.5cm]{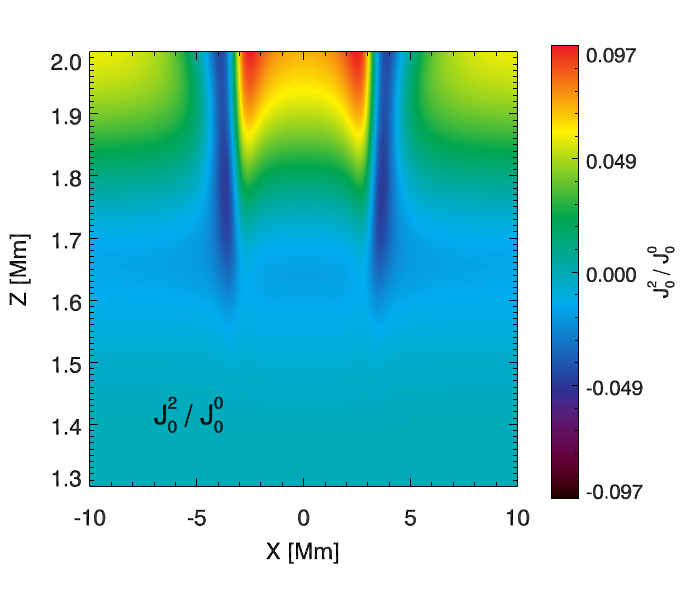} \\
\includegraphics[width=7.5cm]{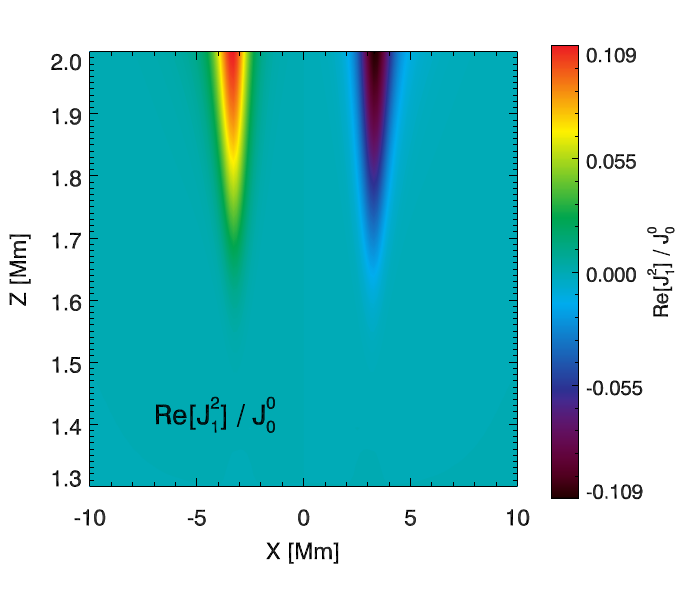} \\
\includegraphics[width=7.5cm]{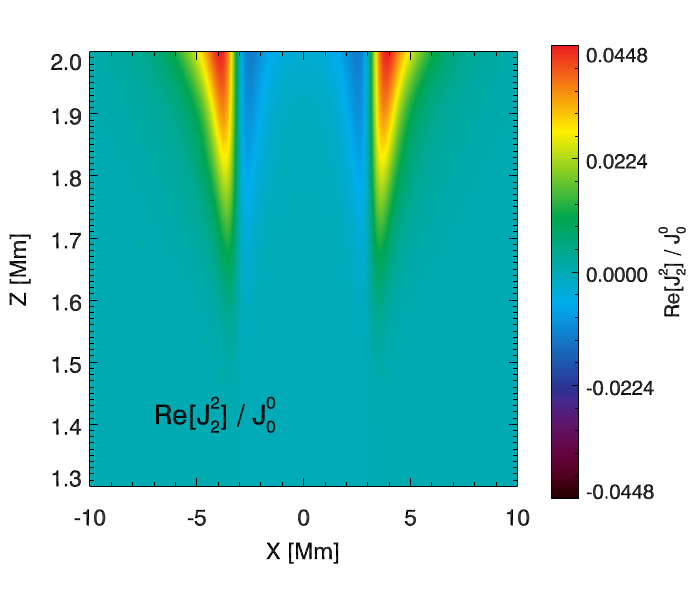} \\
\end{tabular}
\caption[]{Components of the radiation field tensor $J^2_Q/J^0_0$ integrated over the line absorption profile. Top panel: The $J^2_0$ value shows the difference between the amount of vertical and horizontal illumination. Middle and bottom panels: The non-zero $J^2_1$ and $J^2_2$ components at the ribbon edges are due to the anisotropy of horizontal illumination.}
\label{fig:jkq}
\end{figure}

In Fig.\,\ref{fig:jkq}, we show the self-consistent values of the $J^K_Q$ tensors within the model atmosphere. The anisotropy of the line radiation vanishes everywhere below $z\approx 1.5$\,Mm, where the optical thickness of the line becomes significantly larger than unity. Far away from the ribbon, at $|x|\gtrsim 7$\,Mm, the atmosphere can be accurately modelled using the plane-parallel approximation. There we can find the height variation of $J^2_0$ anisotropy which is typical in the isothermal atmospheres \citep[see Fig.\,1 of][]{rms99}.

In accordance with our expectations, we have found that the anisotropy of the radiation is highest at the edges of the ribbon with $|x|\gtrsim 3$\,Mm and $|x|\lesssim 7$\,Mm (see Fig.\,\ref{fig:jkq}), i.e., mainly in the regions of the atmosphere which are not directly affected by the beam itself. All the $J^2_Q$ tensorial components at the height of formation of the line center, i.e., above $z=1.69$\,Mm, are affected by the strong anisotropic emission of the ribbon. It is in these boundary regions where the $J^2_1$ and $J^2_2$ components, which are due to the breaking of the plane-parallel approximation, are nonzero.

Anisotropy in the center of the ribbon is relatively small. This is due to the fact that in the center of the ribbon, the horizontal gradients of physical quantities are smaller than at the ribbon edges. However, in case of a narrow ribbon, one can expect a polarization even in the ribbon center due to different radiation intensities along and perpendicularly to the ribbon orientation.

\begin{figure}[t]
\centering
\begin{tabular}{cc}
\includegraphics[width=7cm]{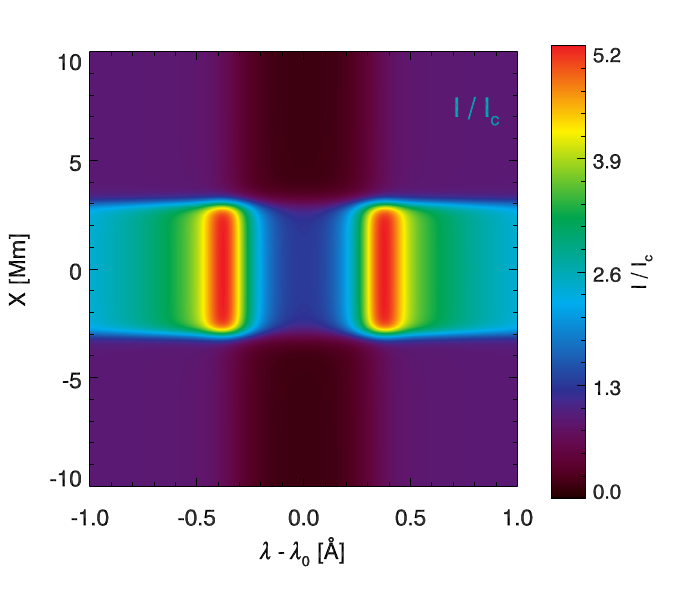} & \includegraphics[width=7cm]{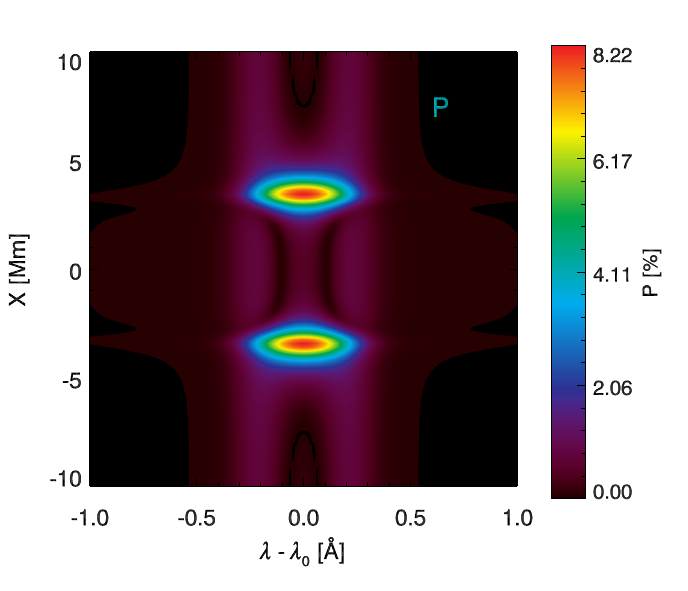} \\
\includegraphics[width=7cm]{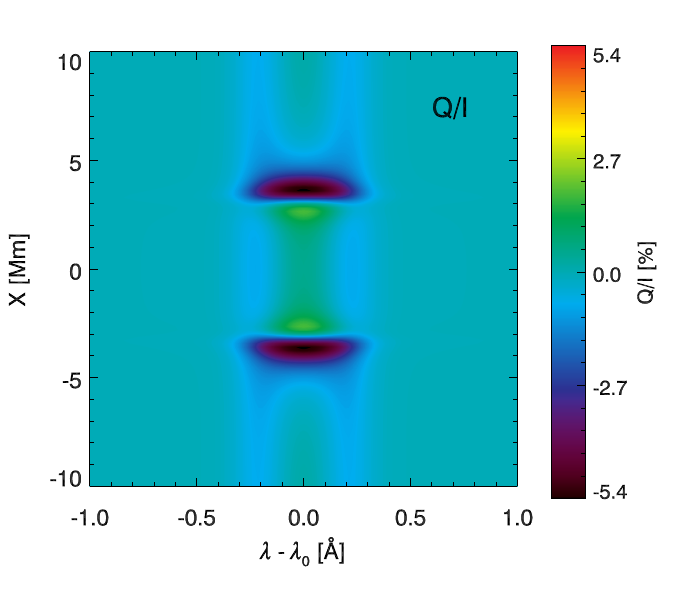} & \includegraphics[width=7cm]{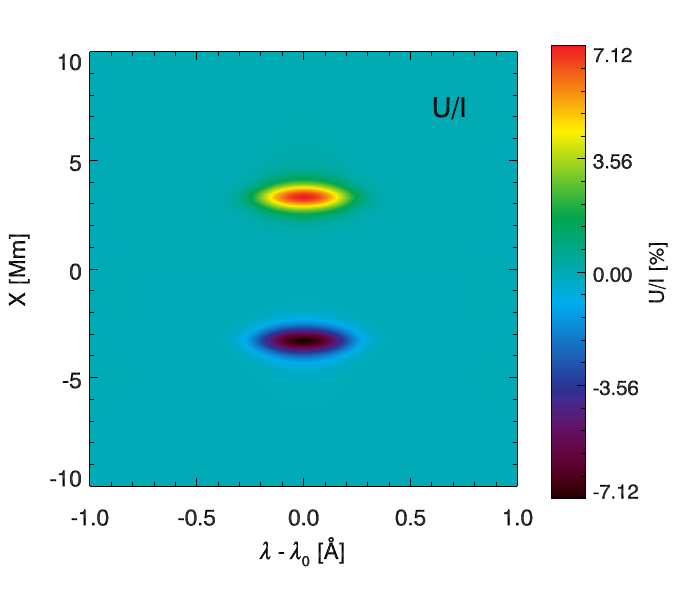} \\
\end{tabular}
\caption[]{
Synthetic spectra of the intensity and linear polarization of the emergent radiation along the spectrograph slit which is parallel to the axis $x$. Azimuth of the observation measured from the positive $x$ axis is $\chi_{\rm LOS}=90^\circ$ and the inclination is $\theta_{\rm LOS}=60^\circ$.
The calculation has been done in the non-magnetized model atmosphere.
Upper left: Specific intensity normalized to the continuum. Note the enhanced emission of the ribbon between $x=-3$ to 3\,Mm, approximately.
Upper right: Total fractional polarization degree $P=\sqrt{Q^2+U^2}/I$.
Bottom left: Fractional polarization $Q/I$ with positive $Q$ reference direction being parallel to the nearest solar limb.
Bottom right: Fractional polarization $U/I$.
}
\label{fig:stokes}
\end{figure}

In Fig.\,\ref{fig:stokes}, we show the emergent Stokes profiles of the line for an inclined line of sight (LOS). The maximum fractional polarization is found at the ribbon edges and the total degree of linear polarization reaches $P\approx 8$\,\%. The degree and orientation of polarization depends on both inclination and azimuth of the LOS, i.e., on the position and orientation of the ribbon on the solar disk. Polarization degree in the central part of the ribbon is significantly reduced with respect to the edges due to the lower radiation field anisotropy in the ribbon center and due to the increased role of the collisional excitation.

\subsection{Magnetized atmosphere and the Hanle effect}

Interestingly enough, in many spectropolarimetric observations, the orientation of polarization vector is either perpendicular or parallel to the nearest solar limb regardless of the ribbon orientation at the solar surface \citep{xu05,henoux13}. This fact is usually used for advocating the role of impact polarization due to the particle beams of various energies propagating along the vertical magnetic field lines. However, the magnetic field itself can directly affect the line polarization via the Hanle effect. We include this mechanism in the calculations presented in this section.

We have kept the model atmosphere and the model atom as in Sect.\,\ref{ssec:nonmag} but we have included a uniform vertical magnetic field of strength $B=500$\,G, i.e., well above the Hanle-effect saturation of our line. Given the fact that the radiation field is not cylindrically symmetric at the ribbon edges, the Hanle effect of the vertical field modifies the atomic polarization \citep[e.g.,][]{rms11}. The atomic coherences $\rho^2_{Q\neq 0}$ are practically removed by the action of the vertical magnetic field in the saturation regime and only the population $\rho^0_0$ and alignment $\rho^2_0$ of the upper atomic level remain.\footnote{Note however, that $J^2_Q$ with $Q\neq 0$ are significantly affected by the presence of magnetic field.}

\begin{figure}[t]
\centering
\begin{tabular}{cc}
\includegraphics[width=7cm]{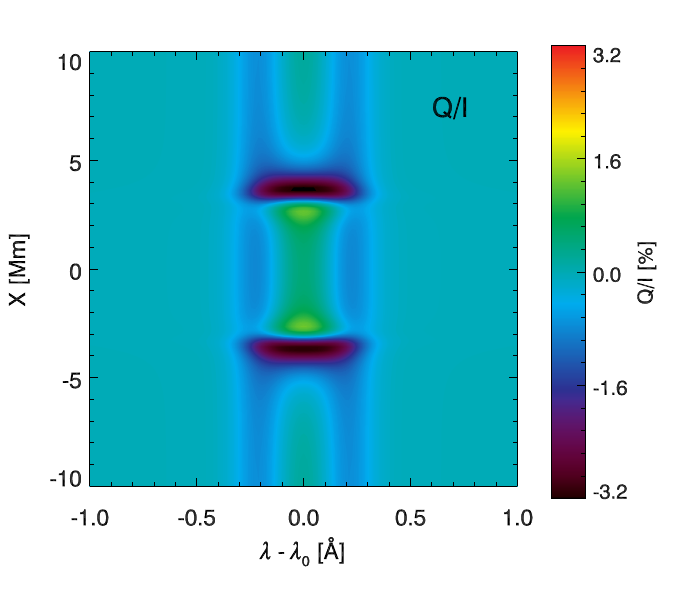} & \includegraphics[width=7cm]{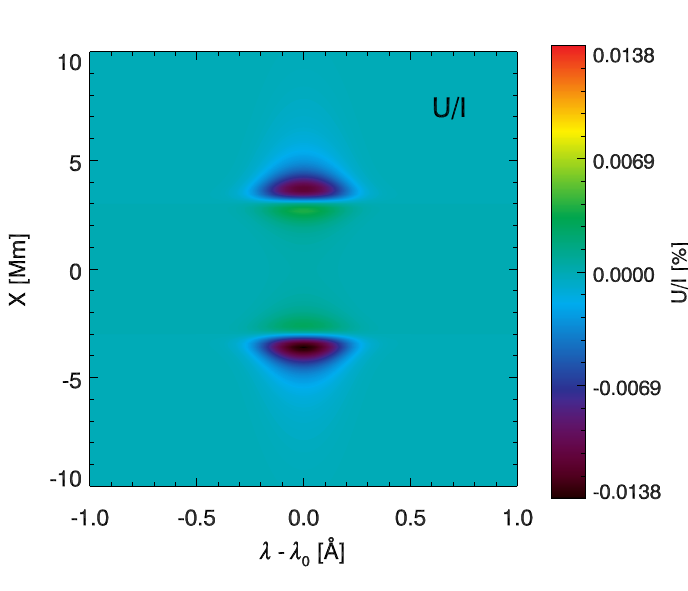}
\end{tabular}
\caption[]{
Emergent linear polarization profiles for the same LOS as in Fig.\,\ref{fig:stokes}, here in the magnetized atmosphere with the uniform vertical magnetic field $B_z=500$\,G. The Stokes $U/I$ parameter is suppressed by the action the Hanle effect and the orientation of the polarization vector is predominantly perpendicular to the nearest solar limb.}
\label{fig:stokesb}
\end{figure}

Similarly to the case of a cylindrically symmetric plane-parallel atmosphere, in which only the $\rho^0_0$ and $\rho^2_0$ components remain, polarization of the emitted photons is either radial or tangential. We demonstrate this in Fig.\,\ref{fig:stokesb} which has been calculated for the same LOS as Fig.\,\ref{fig:stokes} but taking into account the action of the magnetic field. The $U/I$ signal, which quantifies a deviation of the polarization vector from the radial and tangential directions, is negligible with respect to $Q/I$ in the regions of interest. In case of the LOS considered in our example, the orientation of the polarization is radial.

\section{Concluding comments\label{sec:concl}}

We have demonstrated that the horizontal inhomogeneities of the atmosphere may lead to the creation of significant scattering polarization signatures in spectral lines which are consistent with some of the most common observational findings.

We can summarize our results as follows:
\begin{enumerate}
\item The scattering polarization is largest at the ribbon edges due to scattering of the anisotropic ribbon emission. The degree of such polarization can be of the order of several percent. Sufficient spatial resolution of the observations is necessary for detection of this polarization because the region of enhanced polarization is small.
The non-detection of \citet{bianda05} may be due to the low spatial resolution of the observations which was $10''\times 10''$ \citep[for quantitative discussion see Appendix~A of][]{henoux13}.
\item Polarization in the center of the ribbon is smaller than polarization at the edges due to the lower anisotropy of radiation and due to stronger inelastic collisions.
The role of collisional depolarization in the ribbon center due to collisions with thermal electrons and protons may also play a role in the case of the hydrogen Balmer lines \citep{stepan07} but it is questionable for the lines of other species (Na\,D$_2$, Ca\,{\sc ii}\,K) which are mainly depolarized by collisions with neutral hydrogen whose density actually decreases in the ribbons.
\item In the presence of a strong vertical magnetic field, the relaxation of atomic coherences due to the Hanle effect causes emission of predominantly radial or tangential spectral line polarization, depending the particular spectral line, on the LOS, and on the anisotropy of the radiation field.
\item Polarization can be expected during both the impulsive and gradual phases of a flare. The condition for creation of the polarization is a presence of a localized flare ribbon breaking the plane-parallel symmetry of the atmosphere.
\end{enumerate}

Our model is simple and we have neglected several important facts: We have assumed that the ribbon can be approximated by a 2D structure; we have neglected a possible influence from a different nearby ribbon; we have assumed that the ribbon is symmetric with respect to the $x=0$ plane. The last of these approximations is probably most severe because it ignores the different physical conditions in the inner ribbon edges that are affected by the presence of the post-flare loops.

We have chosen the non-thermal collisional rates so that the resulting emission line intensity profile resembles those of \citet{kasparova02}.
Reduction of the non-thermal collisional rates or reduction of their horizontal gradient leads to decrease of the radiation anisotropy and, consequently, to reduction of the emergent fractional polarization. The scattering polarization signals are further altered by number of physical quantities such as the horizontal gradients of the temperature of the plasmas, the magnetic field direction, collisional depolarizing rates, and the particular spectral line under consideration. Different combinations of these parameters may lead to a large variety of the observed polarization signals. Exploration of such parameter space should be a subject of future detailed investigations.

\acknowledgements
We are grateful to Javier Trujillo Bueno (IAC) and Stanislav Gun\'ar (AIAS) for their valuable comments to the manuscript. Financial support by the Grant Agency of the Czech Republic through grants \mbox{P209/12/P741} and \mbox{P209/12/1652}, and by the project \mbox{RVO:67985815} is gratefully acknowledged.

\newpage

\clearpage 

\clearpage

\clearpage

\clearpage

\end{document}